\documentclass{naturemodified}
\usepackage{graphicx}
\usepackage{amsmath}
\usepackage{amssymb}
\usepackage{epstopdf}
\newcommand{\onlinecite}[1]{\hspace{-1 ex} \nocite{#1}\citenum{#1}}

\newcommand{\bfm}{{\boldsymbol{m}}}

\newcommand{\bfmm}{{\boldsymbol{M}}}
\newcommand{\bfqq}{{\boldsymbol{Q}}}
\newcommand{\bfrr}{{\boldsymbol{R}}}

\title{Hedgehog spin-vortex crystal stabilized in a hole-doped iron-based superconductor}

\author{W. R. Meier$^{1,2}$,
	Q.-P. Ding$^{1,2}$,
	A. Kreyssig$^{1,2}$,
	S. L. Bud'ko$^{1,2}$,
	A. Sapkota$^{1,2}$,
	K. Kothapalli$^{1,2}$,
	V. Borisov$^{3}$,
	R. Valent\'{i}$^{\,3}$,
	C. D. Batista$^{4,5}$,
	P. P. Orth$^{1,2}$,
	R. M. Fernandes$^{6}$,
	A. I. Goldman$^{1,2}$,
	Y. Furukawa$^{1,2}$,	
	A. E. B\"{o}hmer$^{2}$,
	P. C. Canfield$^{1,2}$
	}

\begin{document}
\maketitle \par
\date{\today}

\begin{affiliations}
\item{Department of Physics and Astronomy, Iowa State University, Ames, Iowa 50011, USA}
\item{Ames Laboratory, U.S. DOE, Iowa State University, Ames, Iowa 50011, USA}
\item{Institute of Theoretical Physics, Goethe University Frankfurt am Main, 60438 Frankfurt am Main, Germany}
\item{Department of Physics and Astronomy, University of Tennessee, Knoxville, Tennessee 37996, USA}
\item{Quantum Condensed Matter Division and Shull-Wollan Center,
	Oak Ridge National Laboratory, Oak Ridge, Tennessee 37831, USA}
\item{School of Physics and Astronomy, University of Minnesota, Minneapolis, Minnesota 55455, USA}
\end{affiliations}

Magnetism is widely considered to be a key ingredient of unconventional superconductivity. In contrast to cuprate high-temperature superconductors, antiferromagnetism in Fe-based superconductors (FeSCs) is characterized by a pair of magnetic propagation vectors\cite{Paglione2010_FeBasedSuperconductivity,Dai_2015_FeBasedSCsNeutrons}. Consequently, three different types of magnetic order are possible. Of theses, only stripe-type spin-density wave (SSDW) and spin-charge-density wave (SCDW) orders have been observed\cite{Dai_2015_FeBasedSCsNeutrons,Allred_2016_C4inFeArsenides,Avci_2014_C4inBaNaFe2As2}.
A realization of the proposed spin-vortex crystal (SVC) order is noticeably absent. 
We report a magnetic phase consistent with the hedgehog variation of SVC order in Ni- and Co-doped CaKFe$_{4}$As$_{4}$ based on thermodynamic, transport, structural and local magnetic probes combined with symmetry analysis. 
The exotic SVC phase is stabilized by the reduced symmetry of the CaKFe$_{4}$As$_{4}$ structure.
Our results suggest that the possible magnetic ground states in FeSCs have very similar energies, providing an enlarged configuration space for magnetic fluctuations to promote high-temperature superconductivity.

\begin{center}
	\includegraphics[width=8.6cm]{FeAsMagOrdersFig5.pdf}%
	\label{MagOrders}
\end{center}
\begin{center}Figure 1\end{center}
\noindent {\footnotesize \bf Figure 1. Schematics of possible magnetic order types in CaKFe$_{4}$As$_{4}$.} \footnotesize \textbf{a-d}, Sketches of four magnetic moment patterns on an Fe-As layer in the CaKFe$_{4}$As$_{4}$ structure associated with $\textbf{\textit{Q}}_{\textbf{1}}$\,$=$\,$(\pi,0)$, $\textbf{\textit{Q}}_{\textbf{2}}$\,$=$\,$(0,\pi)$ magnetic propagation vectors. The upper yellow square in \textbf{a} and \textbf{e} represents the projection of the CaKFe$_{4}$As$_{4}$ unit cell. The magnetic unit cells are represented by the central yellow squares in \textbf{a-d}. The brown arrows represent the magnetic moments at the Fe sites and the blue and green arrows the hyperfine field ($\textbf{\textit{H}}_{\mathrm{hf}}$) at the inequivalent As1 and As2 sites. 
In the orthorhombic stripe spin-density wave (SSDW) in \textbf{a} all arsenic atoms experience $\textbf{\textit{H}}_{\mathrm{hf}}$ normal to the plane. 
In the spin-charge-density wave (SCDW), "C$_{4}$ phase\cite{Avci_2014_C4inBaNaFe2As2}", in \textbf{b} every other Fe site supports a moment normal to the plane and the As sites experience an in-plane $\textbf{\textit{H}}_{\mathrm{hf}}$. 
In the hedgehog spin-vortex crystal (SVC) phase in \textbf{c} Fe moments display an "all in" or "all out" arrangement centered around the As1 sites generating an out-of-plane $\textbf{\textit{H}}_{\mathrm{hf}}$\,$\ne$\,$\textbf{0}$ at these sites and $\textbf{\textit{H}}_{\mathrm{hf}}$\,$=$\,$\textbf{0}$ at the As2 sites.
In the loops SVC in \textbf{d} the magnetic moments form loops around As1 and neither As site experiences a $\textbf{\textit{H}}_{\mathrm{hf}}$. 
\textbf{e}, The chemical structure of CaKFe$_{4}$As$_{4}$. Note the inequivalent As1 and As2 sites adjacent to K and Ca planes, respectively.
\textbf{f}, Section of the Fe-As plane with a hedgehog SVC moment arrangement. Spin up currents between the iron atoms, $\textbf{\textit{J}}_{\mathrm{s}}$ (yellow arrows), generate an electric field, \textbf{\textit{E}} (red arrows), which couples to asymmetric shifts of the two arsenic sites. Unlike in the CaFe$_{2}$As$_{2}$ structure, an asymmetric arrangement of arsenic atoms is imposed by the crystallographic symmetry in CaKFe$_{4}$As$_{4}$ providing a symmetry-breaking field that favors the SVC-type phases.

\normalsize
The many families of FeSCs provide a diverse platform for investigating the fundamental nature and applications of high-temperature superconductivity\cite{Paglione2010_FeBasedSuperconductivity,Pallecchi_2015_ApplicationsOfFeBasedSCs}. 
In these compounds, magnetic order and superconductivity seem intertwined and the ground-state can be tuned by pressure and doping. The FeSCs share a common chemical motif; stacked Fe-pnictide or chalcogenide layers. 
Magnetism in FeSCs is characterized by antiferromagnetic correlations with two symmetry-equivalent propagation vectors, $\textbf{\textit{Q}}_{\textbf{1}}$\,$=$\,$(\pi,0)$ and $\textbf{\textit{Q}}_{\textbf{2}}$\,$=$\,$(0,\pi)$ (using the single-iron Brillouin zone notation)\cite{Paglione2010_FeBasedSuperconductivity,Dai_2015_FeBasedSCsNeutrons}.
Magnetic order with these propagation vectors can be described by the spatial variation of the magnetic moments at the iron sites at positions \textbf{\textit{R}} (Refs.~\onlinecite{Lorenzana_2008_CompetingOrdersInFeSCs,Eremin_2010_SDWsInFeSCs,Brydon_2011_GinzburLandauFeSCs,Cvetkovic_2013_SpGrpSymmAndSOHamltnFeSCers,Wang_2015_TetragMagOrderFeArsenides,Christensen_2015_SpinReorientation,Gastiasoro_2015_2QinFeSCs,Hoyer_2016_DisorderPromotedC4Phase,Fernandes_2016_VestigialOrders,OHalloran_2017_SVCinFeSCers}),
\begin{equation}
\textbf{\textit{m}}(\textbf{\textit{R}})=\textbf{\textit{M}}_{\textbf{1}} \cos{\left(\textbf{\textit{Q}}_{\textbf{1}}\cdot\textbf{\textit{R}}\right)}+\textbf{\textit{M}}_{\textbf{2}}\cos{\left(\textbf{\textit{Q}}_{\textbf{2}}\cdot\textbf{\textit{R}}\right)}.
\label{moments}
\end{equation} 
Three types of magnetic order (Fig.~1a-d) are defined by the relationship between the magnetic order parameters $\textbf{\textit{M}}_{\textbf{1}}$ and $\textbf{\textit{M}}_{\textbf{2}}$. The orthorhombic SSDW order (Fig.~1a) has only one nonzero $\textbf{\textit{M}}_{\textbf{\textit{i}}}$. A SCDW (Fig.~1b) is a superposition of the two contributions with $\textbf{\textit{M}}_{\textbf{1}}$\,$=$\,$\pm$$\textbf{\textit{M}}_{\textbf{2}}$. 
Finally, tetragonal SVC orders (Fig.~1c,d) are defined by $|\textbf{\textit{M}}_{\textbf{1}}|$\,$=$\,$|\textbf{\textit{M}}_{\textbf{2}}|$ with non-collinear vectors $\textbf{\textit{M}}_{\textbf{1}}$\,$\perp$\,$\textbf{\textit{M}}_{\textbf{2}}$, both in the Fe-plane. 

The existence of two vector order parameters, $\textbf{\textit{M}}_{\textbf{1}}$ and $\textbf{\textit{M}}_{\textbf{2}}$, potentially provides a large configuration space for magnetic fluctuations. 
They are maximized when different types of magnetic order are close in energy, which may be key to the high-temperature superconductivity in the FeSCs. This makes FeSCs exceptional, as most unconventional superconductors support only a single magnetic order parameter.
Although the majority of iron-based superconductors exhibit only SSDW magnetic order, several hole-doped \textit{Ae}Fe$_{2}$As$_{2}$ (\textit{Ae}\,=\,Ca, Sr, Ba) compounds demonstrate a transition from SSDW to SCDW for compositions near optimal superconductivity\cite{Avci_2014_C4inBaNaFe2As2,Bohmer_2015_C4inBaKFe2As2,Allred_2016_C4inFeArsenides,Hassinger_2016_C4inBaKFe2As2Pressure,Mallett_2015_muSRonC4inBaKFe2As2}.
An experimental realization of SVC order would demonstrate the diversity of magnetic ground-states in FeSCs, but has not been reported to date. 

Application of a suitable symmetry-breaking field may coax the system to condense into one of the three types of magnetic orders as they break distinct symmetries\cite{Fernandes_2016_VestigialOrders}. 
For example, in-plane uniaxial strain favors SSDW as it breaks tetragonal symmetry in the same way and has been widely employed to study this order\cite{Chu_2012_FeBasedSCsElastoresistivity,Blomberg_2011_DetwinnedResistivitySrFe2As2}. It is difficult to imagine appropriate, externally applicable symmetry-breaking fields for SCDW or SVC orders. A SVC phase breaks the glide symmetry across the Fe-As planes\cite{Fernandes_2016_VestigialOrders}, which is present in most FeSCs. 
Consequently, breaking this glide symmetry could favor SVC order\cite{OHalloran_2017_SVCinFeSCers}.
In the recently discovered stoichiometric \textit{AeA}Fe$_{4}$As$_{4}$ (\textit{Ae}\,=\,Ca, Sr; \textit{A}\,=\,K, Rb, Cs) superconductors\cite{Iyo2016_AeAFe4As4}, the alternating \textit{Ae} and \textit{A} atom planes inherently break this glide symmetry resulting in two inequivalent As sites (Fig. 1e). 
Consequently, these compounds could provide a unique opportunity to realize and study a SVC phase. Unfortunately, the parent \textit{AeA}Fe$_{4}$As$_{4}$ compounds do not order magnetically\cite{Iyo2016_AeAFe4As4,Meier2016_CaKFe4As4}. 
Electron count and experimental properties suggests that CaKFe$_4$As$_4$ can be considered analogous to (Ba$_{0.5}$K$_{0.5}$)Fe$_{2}$As$_{2}$ (Ref.~\onlinecite{Meier2016_CaKFe4As4}).
In the latter compound, the SSDW is suppressed by hole doping\cite{Paglione2010_FeBasedSuperconductivity} (substituting K into BaFe$_2$As$_2$). Inspired by this analogy\cite{Zinth_2011_BaKFeCo2As2}, magnetic order could be induced by electron doping CaKFe$_4$As$_4$.
Here, we report that adding electrons via substitution of Co or Ni for Fe in CaKFe$_{4}$As$_{4}$ does induce antiferromagnetism consistent with the hedgehog variation of SVC. 
We stabilize this phase using the chemical structure as a symmetry-breaking field illustrating the multiple competing, nearly-degenerate magnetic phases in FeSCs. This competition may be an important contributor to high-temperature superconductivity as enhanced magnetic fluctuations can boost pairing. 

\begin{center}
	\includegraphics[width=8.6cm]{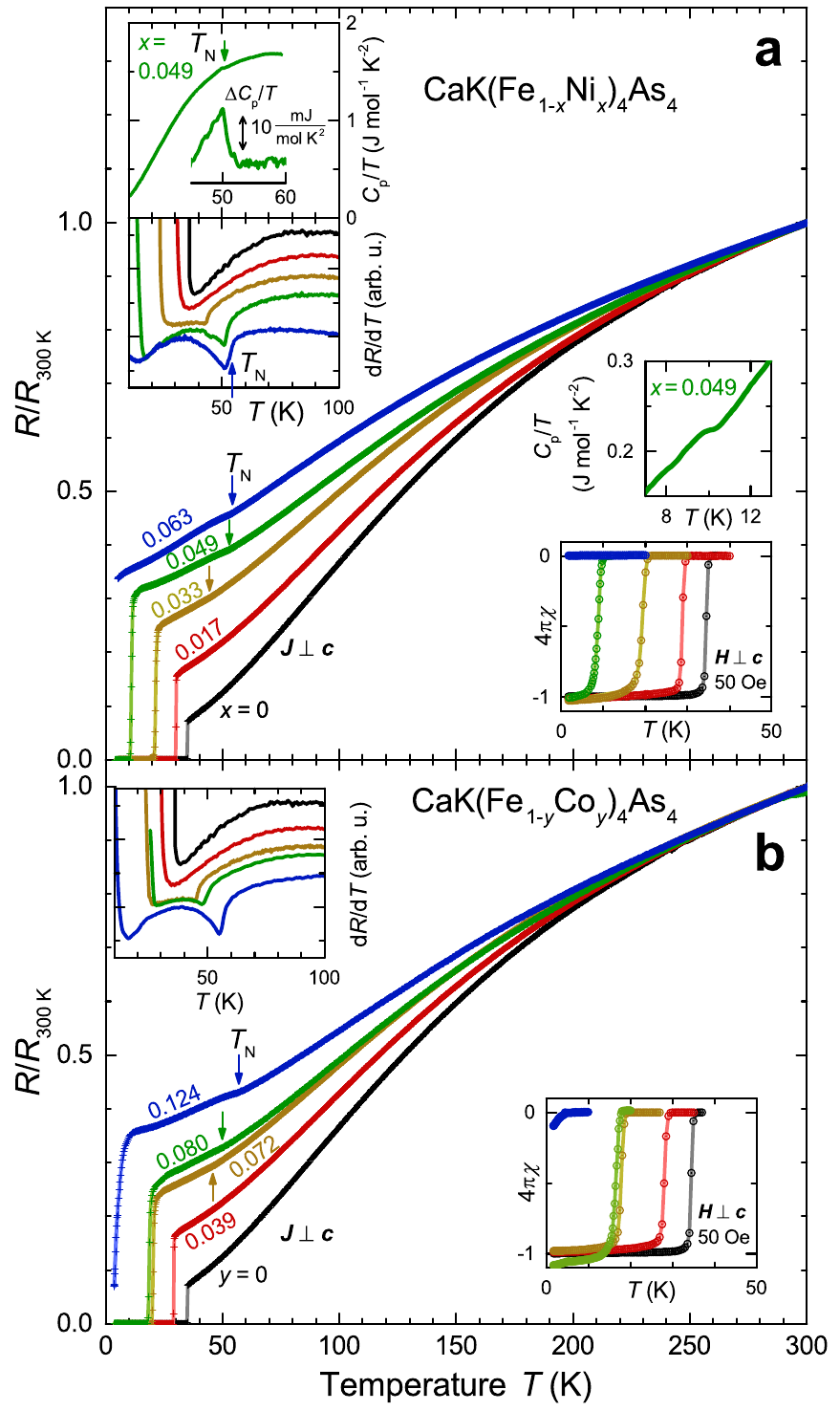}
	\label{RT_4PiChi}
\end{center}
\begin{center}Figure 2\end{center}
\noindent {\footnotesize \bf Figure 2. Resistance, magnetic susceptibility, and heat capacity of Co- and Ni-doped CaKFe$_{4}$As$_{4}$.} \footnotesize \textbf{a}, Temperature dependence of normalized resistance, $R/R_{300\,\mathrm{K}}$, of CaK(Fe$_{1-\textit{x}}$Ni$_{\textit{x}}$)$_{4}$As$_{4}$ single crystals demonstrating the suppression of the superconducting transition temperature, $T_{\mathrm{c}}$, and the emergence of a kink-like feature at $T_{\mathrm{N}}$ with increasing Ni content, $x$. The anomaly at $T_{\mathrm{N}}$ appears clearly as a step in the derivative of resistance, d\textit{R}/d\textit{T}, and specific-heat, $\textit{C}_{p}$, of $x$\,$=$\,$0.049$ (upper left insets). In the magnified portion in the specific-heat inset, data from a $x$\,$=$\,$0.017$ sample was subtracted to accentuate the step-like anomaly characteristic of a bulk, second-order transition at $T_{\mathrm{N}}$. The zero-field cooled (ZFC) volumetric magnetic susceptibility, $4\pi\chi$, (bottom right inset) demonstrates the suppression of $T_{\mathrm{c}}$ with increasing $x$ and full superconducting shielding for $x$\,$\leq$\,$0.049$. The upper right inset shows a magnified view of $\textit{C}_{p}/T$ of the $x$\,$=$\,$	0.049$ sample around $T_{\mathrm{c}}$. 
\textbf{b}, Evolution of resistance and ZFC magnetic susceptibility in CaK(Fe$_{1-\textit{y}}$Co$_{\textit{y}}$)$_{4}$As$_{4}$. $T_{\mathrm{c}}$ is suppressed by Co substitution and the kink-like feature at $T_{\mathrm{N}}$ emerges in the resistance data similar to the effect of Ni substitution. 

\begin{center}
	\includegraphics[width=8.6cm]{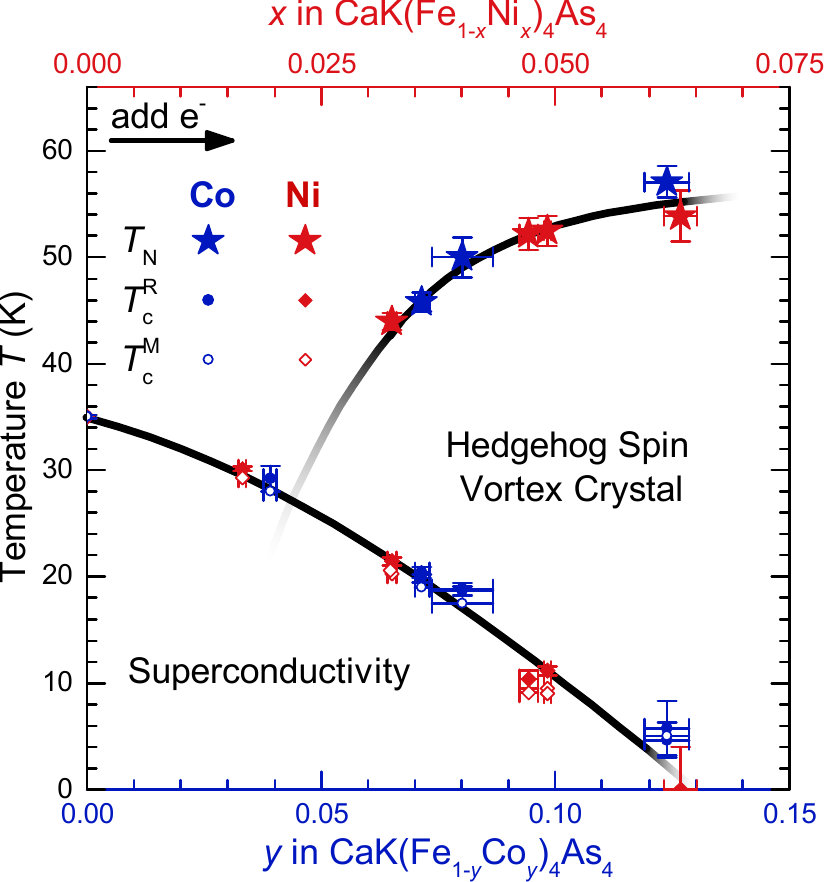}%
	\label{PhaseDiagram}
\end{center}
\begin{center}Figure 3\end{center}
\noindent {\footnotesize \bf Figure 3. Common phase diagram of Co- and Ni-doped CaKFe$_{4}$As$_{4}$.} \footnotesize Doping CaKFe$_{4}$As$_{4}$ with either Co or Ni suppresses the superconducting transition temperature, $T_{\mathrm{c}}$, and stabilizes a hedgehog spin-vortex crystal below $T_{\mathrm{N}}$. $T_{\mathrm{c}}^R$ and $T_{\mathrm{c}}^M$ were determined by resistance and magnetization measurements, respectively. The Ni-concentration, $x$ on the upper axis, is scaled by a factor of two with respect to the Co-concentration, $y$, which maps the transition temperatures of the two series onto each other. This scaling is consistent with an electron doping picture where Ni substitution adds twice as many electrons per atom as Co. Transition temperatures of CaK(Fe$_{1-\textit{x}}$Ni$_{\textit{x}}$)$_{4}$As$_{4}$ determined by heat capacity, M\"{o}ssbauer spectroscopy, and nuclear magnetic resonance are consistent with this phase diagram. Horizontal error bars represent the standard deviation of 12 composition points for each sample. Vertical error bars indicate transition width.

\normalsize
Figure 2a presents the electrical resistance of CaK(Fe$_{1-\textit{x}}$Ni$_{\textit{x}}$)$_{4}$As$_{4}$. With increasing nickel content, a kink at temperature $T_{\mathrm{N}}$ appears and rises to $\sim$$55$\,K as the superconducting critical temperature, $T_{\mathrm{c}}$, is suppressed. The resistance of CaK(Fe$_{1-\textit{y}}$Co$_{\textit{y}}$)$_{4}$As$_{4}$ (Fig.~2b) exhibits a similar trend. The heat capacity of CaK(Fe$_{0.951}$Ni$_{0.049}$)$_{4}$As$_{4}$ (insets into Fig.~2a) reveals clear signatures of second-order phase transitions at $T_{\mathrm{N}}$ and $T_{\mathrm{c}}$. The transition temperatures inferred from these data are plotted in Fig.~3, demonstrating that the phase diagrams of Co- and Ni-doped CaKFe$_{4}$As$_{4}$ map onto each other if the Ni fraction is scaled by two. This is consistent with an electron doping picture where substituting Ni for Fe supplies twice as many electrons as Co (Ref.~\onlinecite{Canfield2010_TMdopedBaFe2As2}).

\begin{center}
	\includegraphics[width=17cm]{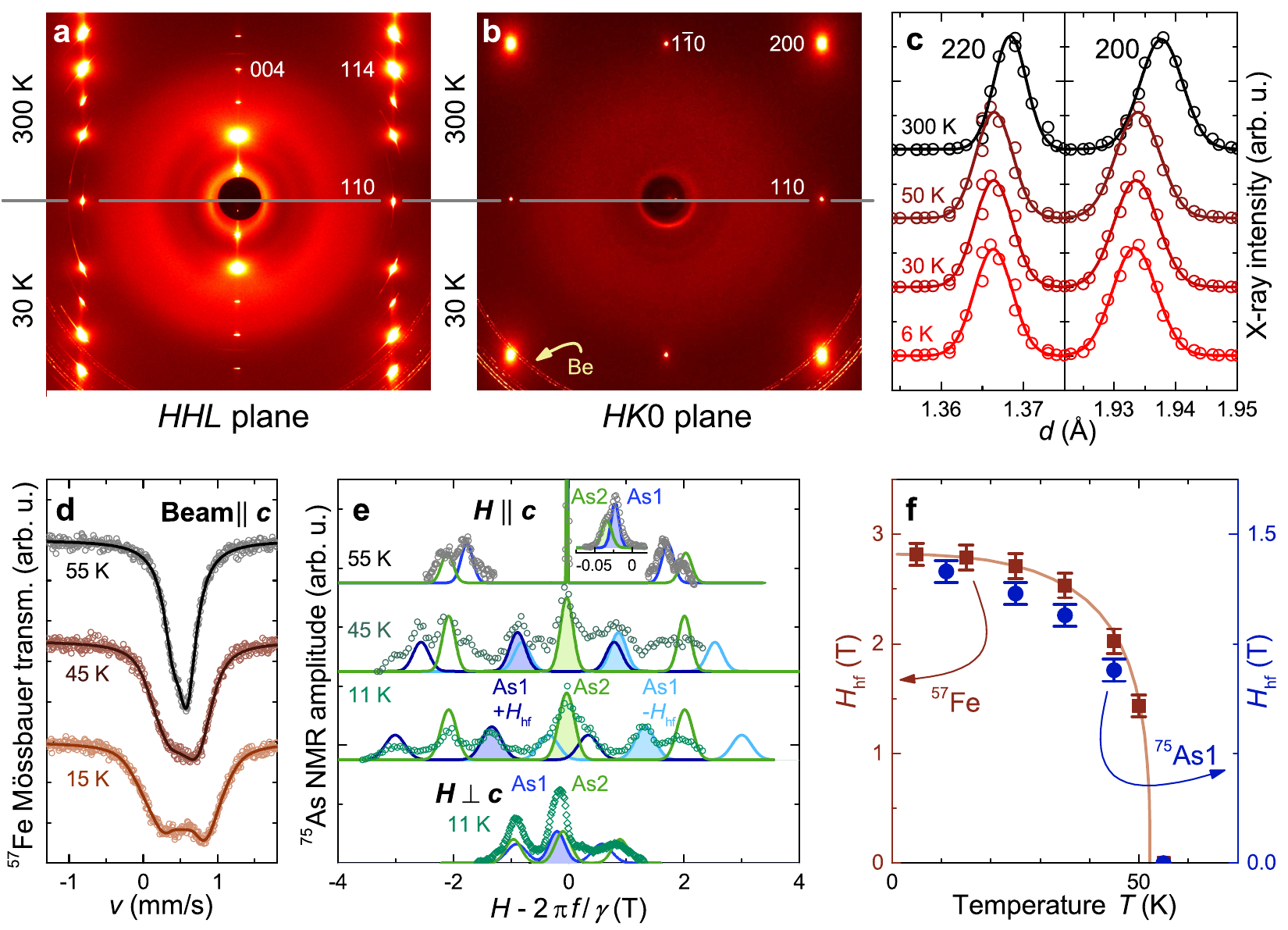}%
	\label{MicroscopicData}
\end{center}
\begin{center}Figure 4\end{center}
\noindent {\footnotesize \bf Figure 4. Experimental insights into the nature of the magnetic phase of CaK(Fe$_{0.951}$Ni$_{0.049}$)$_{4}$As$_{4}$ below $T_{\mathrm{N}}=$~53\,K.} 
\footnotesize \textbf{a},\textbf{b}, Single crystal $HHL$ and $HK0$ diffraction patterns at $300$\,K (upper halves) and $30$\,K (lower halves). No additional diffraction peaks appear at $30$\,K, counter-indicating a chemical superlattice. The Be sample enclosure produces diffraction rings. \textbf{c}, 220 and 200 diffraction peaks (integrated azimuthally) vs. plane spacing $d$. 
The peak shapes remain unchanged below $T_{\mathrm{N}}$ placing an upper relative limit of $4$$\times$$10^{-4}$ on lattice distortions in either direction.
\textbf{d}, $^{57}$Fe M\"{o}ssbauer spectra on a mosaic of $34$ single crystals. Broadening of the absorption peak on cooling through $T_\mathrm{N}$ indicates a hyperfine field, $\textbf{\textit{H}}_{\mathrm{hf}}$, produced by an ordered moment at the Fe-sites. The ratios of the peaks in the fitted sextet (solid lines) reveal that these ordered moments lie perpendicular to the crystallographic $c$-axis. 
\textbf{e}, Field-swept $^{75}$As (gyromagnetic ratio, $\gamma/2\pi$\,$=$\,$7.2919$\,MHz/T) nuclear magnetic resonance (NMR) spectra obtained at $f$\,$=$\,$43.2$\,MHz with applied magnetic field, $\textbf{\textit{H}}$, along the crystallographic $c$-axis (upper three spectra) and with $\textbf{\textit{H}}$\,$\perp$\,$\textbf{\textit{c}}$ (bottom spectrum). Simulated peaks from the two inequivalent As1- and As2-sites are denoted by hues of blue and green, respectively. Above $T_{\mathrm{N}}$ each As site produces a central peak (shaded area) and a pair of electric quadrapole satellites. With $\textbf{\textit{H}}$\,$\parallel$\,$\textbf{\textit{c}}$, the As1 triplet splits into two triplets on cooling through $T_{\mathrm{N}}$ indicating presence of a $\textbf{\textit{H}}_{\mathrm{hf}}$ component parallel to \textbf{\textit{c}} at half of the As1-sites and antiparallel to \textbf{\textit{c}} at the other half. The As2 triplet remains un-split. With $\textbf{\textit{H}}$\,$\perp$\,$\textbf{\textit{c}}$, neither triplet is split at 11\,K indicating an absence of in-plane hyperfine fields at either As site. 
\textbf{f}, The temperature dependence of $|\textbf{\textit{H}}_{\mathrm{hf}}|$ at the Fe- (squares, from M\"{o}ssbauer) and As1-sites (circles, from NMR), consistent with a second-order magnetic phase transition at $T_{\mathrm{N}}$. The error bars represent standard fitting error ($^{57}$Fe) and estimated uncertainty ($^{75}$As).

\normalsize
Consistent observation of a transition at $T_{\mathrm{N}}$ in both doping series calls for investigation of its nature. Figures 4a-c show high-energy x-ray diffraction results from CaK(Fe$_{0.951}$Ni$_{0.049}$)$_{4}$As$_{4}$. The peak shapes (Fig. 4c) are unchanged below $T_{\mathrm{N}}$\,$=$\,$53$\,K indicating an absence of lattice distortions.
No superlattice peaks appear in the low-temperature diffraction patterns in the lower halves of Fig.~4a,b.
These results suggest that the geometry and multiplicity of the chemical unit cell are unchanged.
$^{57}$Fe M\"{o}ssbauer spectroscopy on crystals of CaK(Fe$_{0.951}$Ni$_{0.049}$)$_{4}$As$_{4}$ in Fig.~4d reveals clear structure in the spectra below $55$\,K demonstrating a hyperfine field, $\textbf{\textit{H}}_{\mathrm{hf}}$, at the Fe-sites, i.e., magnetic order. 
A fit of the spectrum at base temperature suggests a $H_{\mathrm{hf}}$ distribution centered around $3$\,T (corresponding to $\sim$$0.2$\,$\mathrm{\mu}_\mathrm{B}$ ordered moment\cite{Rotter2008_SDWinBaFe2As2}) and in-plane orientation of the iron moments.
The temperature dependence of $H_{\mathrm{hf}}$ (Fig.~4f) is characteristic of a second-order phase transition, consistent with heat capacity.
$^{75}$As nuclear magnetic resonance (NMR) is employed to probe the $\textbf{\textit{H}}_{\mathrm{hf}}$ at the inequivalent As1 and As2 sites of the CaKFe$_{4}$As$_{4}$ structure. Figure 4e reveals a finite $\textbf{\textit{H}}_{\mathrm{hf}}$\,$\parallel$\,$\textbf{\textit{c}}$ at As1 (close to K) and $\textbf{\textit{H}}_{\mathrm{hf}}=\textbf{0}$ at As2. The temperature dependence of $H_{\mathrm{hf}}$ at As1 mirrors that at the Fe site (Fig.~4f). The assignment of As1 and As2 is made following Ref. \onlinecite{Cui2017_CaKFe4As4NMR}.

The M\"{o}ssbauer and NMR spectra are inconsistent with the hyperfine fields at Fe and As in SSDW and SCDW phases (see Fig.~1a,b and Supplemental Material\cite{Supplement}).
Specifically, the M\"{o}ssbauer spectra cannot be described by the moment motif of the SCDW phase where half the Fe-sites would experience $H_{\mathrm{hf}}$\,=\,$0$.
Furthermore, the diffraction data in Fig.~4 are inconsistent with an orthorhombic distortion. Of the possibilities depicted in Fig.~1, only SVC-type order remains. 
To systematically consider the possible magnetic phases of CaK(Fe$_{0.951}$Ni$_{0.049}$)$_{4}$As$_{4}$, a representation analysis of CaKFe$_{4}$As$_{4}$'s magnetic space group, \textit{P4/mmm1'}, was performed, similar to Ref.~\onlinecite{Khalyavin_2014}. There are $96$ commensurate magnetic structures allowed for the chemical structure\cite{Iyo2016_AeAFe4As4} of CaKFe$_{4}$As$_{4}$. Possible incommensurate magnetic phases are discussed in supplemental information\cite{Supplement}. Only $24$ structures retain the original tetragonal chemical unit cell as suggested by the diffraction results. All of these are accessible by a second-order phase transition\cite{Supplement} and have one common magnitude for all iron moments. However, the ordered Fe moments in $8$ of these (including ferromagnetism and N\'{e}el order) are normal to the Fe-planes, inconsistent with the M\"ossbauer results. 
The remaining $16$ structures are all modifications of SVC order. 
There are two motif variations\cite{OHalloran_2017_SVCinFeSCers}, "hedgehog" with $\textbf{\textit{M}}_{\textbf{\textit{i}}}$\,$\parallel$\,$\textbf{\textit{Q}}_{\textbf{\textit{i}}}$ (Fig.~1c)
and "loops" with $\textbf{\textit{M}}_{\textbf{\textit{i}}}$\,$\perp$\,$\textbf{\textit{Q}}_{\textbf{\textit{i}}}$ (Fig.~1d).
Each variation can be centered on either As1 or As2 and can have four different stacking patterns along the crystallographic $c$-axis\cite{Supplement}. 
Only the hedgehog SVC, with the centering depicted in Fig. 1c, is consistent with the $\textbf{\textit{H}}_{\mathrm{hf}}$ at As1 observed by NMR.

We argue that this peculiar, non-collinear magnetic phase is stabilized by the specific crystal structure of CaKFe$_{4}$As$_{4}$. The inequivalence of the staggered As sites (Fig. 1e) generates a symmetry-breaking field that couples to the spin vorticity, $\textbf{\textit{M}}_{\textbf{1}} \times \textbf{\textit{M}}_{\textbf{2}}$, characteristic of SVC order.
One possible mechanism is the electrostatic coupling of the As atoms to spin current loops associated with the magnetic moment motif\cite{Fernandes_2016_VestigialOrders,Supplement} (Fig.~1f). These spin currents generate an electric field analogous to the magnetic field produced by an electric current\cite{Sun_2004_SpinCurrentEField}. The resulting electrostatic force couples the asymmetric shifts of the As atoms to the spin currents.
The preference for the hedgehog SVC variation over loops variation is likely a consequence of spin-orbit coupling, which generally favors $\textbf{\textit{M}}_{\textbf{\textit{i}}}$\,$\parallel$\,$\textbf{\textit{Q}}_{\textbf{\textit{i}}}$, observed experimentally in the case of the SSDW order of the parent \textit{Ae}Fe$_2$As$_2$ compounds\cite{Dai_2015_FeBasedSCsNeutrons}, and theoretically in Ref. \onlinecite{Christensen_2015_SpinReorientation}. 

Our identification of a hedgehog SVC order in Co- and Ni-doped CaKFe$_{4}$As$_{4}$ sheds new light on the curious landscape of magnetic phases in FeSCs arising from two magnetic propagation vectors. 
Previous results indicate that SSDW and SCDW phases have comparable energies in some hole-doped FeSCs\cite{Allred_2016_C4inFeArsenides,Avci_2014_C4inBaNaFe2As2,Bohmer_2015_C4inBaKFe2As2,Hassinger_2016_C4inBaKFe2As2Pressure}. 
Our observation in effectively hole-doped CaKFe$_{4}$As$_{4}$ reveals that SVC order is competitive as well and demonstrates the sensitivity of magnetism in FeSCs. 
This phenomenon is confirmed by first-principle calculations, which can compare the energies of different magnetic configurations.
Previous calculations needed to invoke a SVC-like order to reproduce the structural collapse of CaKFe$_{4}$As$_{4}$ under pressure\cite{Kaluarachchi_2017_CaKFe4As4CollapesedTetrag}. 
Our calculations reveal that SSDW and hedgehog SVC are nearly degenerate for pure CaKFe$_{4}$As$_{4}$ but the latter prevails upon electron doping\cite{Supplement}.
The near-degeneracy of these magnetic phases provides an enlarged configuration space for magnetic fluctuations which may have profound implications for magnetically-mediated superconducting pairing\cite{Hirschfeld_2011_review,Chubukov2012_PairingInFeSuperconductors}.

In conclusion, we confidently identified a hedgehog spin-vortex crystal in Co- and Ni-doped CaKFe$_{4}$As$_{4}$. 
The chemical structure of CaKFe$_{4}$As$_{4}$ and spin-orbit coupling conspire to stabilize this phase exposing the complex magnetism in FeSCs.
This first experimental realization of SVC order in a magnetic superconductor opens a new platform to study multi-component, non-collinear magnetic phases and their interplay with superconductivity.

\begin{methods}

Single crystals of Co- and Ni-doped CaKFe$_{4}$As$_{4}$ were grown out of a high-temperature solution rich in transition-metals and arsenic similar to the procedure used for the pure compound\cite{Meier2016_CaKFe4As4,Meier2017_OptimizeCaKFe4As4}. Transition-metal arsenides precursors, $M_{0.512}$As$_{0.488}$ ($M$\,=\,Fe, Co, Ni), were synthesized from ground arsenic lumps (Alfa Aesar $99.9999$\%) and transition-metal powders [Fe (Alfa Aesar $99.9$+\% metals basis), Co (Alfa Aesar $99.8$\% metals basis), Ni (Alfa Aesar $99.9$\% metals basis)] in a rotating fused-silica ampoule in a purpose-built tube furnace as described in Ref.~\onlinecite{Ran2014Thesis}. The iron and cobalt reactions were held at $565^{\circ}$C for $900$\,min then $900^{\circ}$C for $600$\,min before the furnace was switched off. The nickel version was held at $720^{\circ}$C for $600$\,min instead of $900^{\circ}$C because nickel arsenides melt at lower temperatures. Ground transition-metal arsenide precursors $M_{0.512}$As$_{0.488}$, potassium metal globs (Alfa Aesar $99.95$\%), and distilled calcium metal pieces (Ames Laboratory, Materials Preparation Center (MPC) $99.9$\%) were loaded into alumina Canfield Crucible Sets (LSP Industrial Ceramics, Inc.)\cite{Canfield2016_FritDiskCrucibles}. Batches were about $2$\,grams of material with a molar ratio of K\,:\,Ca\,:\,Fe$_{0.512}$As$_{0.488}$\,:\,$M_{0.512}$As$_{0.488}$\,$=$\,$1.2$\,$:$\,$0.8$\,$:$\,$20\,(1-x)$\,$:$\,$20\,x$. This crucible set was sealed in a tantalum metal tube and then in an argon-filled, fused-silica ampoule. These ampoules were heated in a box furnace using the optimum profile described in Ref.~\onlinecite{Meier2017_OptimizeCaKFe4As4}.
The final temperature varied from $920^{\circ}$C to $933^{\circ}$C depending on how the transition metal ratio shifted the solidification temperature. After one to two hours at this final temperature the growth assembly was removed from the furnace, inverted into a centrifuge, and spun to decant the remaining liquid off the crystals\cite{Canfield2016_FritDiskCrucibles}. The metallic plate-like crystals obtained were smaller than those of undoped CaKFe$_4$As$_4$ and cleaved easily along the (001) planes.

Chemical analysis of freshly-cleaved crystal surfaces was performed by wavelength-dispersive x-ray spectroscopy (WDS) on a JEOL JXA-8200 microprobe system. Twelve points were analyzed for each sample. CaWO$_4$, KAlSi$_3$O$_8$, Co metal, Ni metal and FeAs were used as standards. 

Electrical resistance was measured in a standard four-contact geometry with Pt-wires bonded to the samples with silver paint. A Lakeshore Model 370 AC resistance bridge was employed and the temperature environment was provided by a Janis Research SHI-950T 4 Kelvin closed-cycle refrigerator.

Magnetization measurements were performed in a Quantum Design, Magnetic Property Measurement System in a clear plastic straw. The samples were zero-field cooled before applying a magnetic field of $50$\,Oe parallel to the plate-shaped crystals.

The heat capacity data of the samples was measured using a hybrid adiabatic relaxation technique of the heat capacity option in a Quantum Design, Physical Properties Measurement System.

High-energy x-ray diffraction measurements were performed on the six-circle diffractometer at end station 6-ID-D at the Advanced Photon Source, Argonne National Laboratory, using an x-ray energy of $E$\,$=$\,$100.34$\,keV and a beam size of $100\times100$\,$\mu$m$^2$. The CaK(Fe$_{0.951}$Ni$_{0.049}$)$_{4}$As$_{4}$ single-crystal sample was mounted on the cold finger of a He closed-cycle refrigerator. The sample was enclosed by a Be dome and He exchange gas was used to ensure the thermal equilibrium. Diffraction patterns were recorded using a MAR345 image plate detector positioned at $1.474$\,m from the sample. The distance was determined from measurement of powder patterns of CeO$_2$ standard from the National Institute of Standards and Technology. The detector was operated with a pixel size of $100\times100$\,$\mu$m$^2$, and patterns were recorded while rocking the sample through two independent angles up to $\pm$$2.8^{\circ}$ about the axes perpendicular to the incident beam.

$^{57}$Fe M\"{o}ssbauer spectroscopy measurements were performed in a SEE Co.~conventional, constant acceleration type spectrometer in transmission geometry. The $^{57}$Co in Rh source was maintained at room temperature. $34$ cleaved crystals of CaK(Fe$_{0.951}$Ni$_{0.049}$)$_{4}$As$_{4}$ were fixed to a paper disk with Apiezon N grease. An effort was made to keep gaps between crystals to a minimum and the part of the disk not covered by crystals was coated with tungsten powder (Alfa Aesar $99.9$\% metals basis). This mosaic was positioned so that the gamma ray beam was parallel to the crystallographic \textit{c}-axes. The sample temperature was maintained using a Janis SHI-850-5 closed cycle refrigerator with vibration damping. The driver velocity was calibrated using an $\alpha$-Fe foil. The M\"ossbauer spectra were fitted using the commercial software package MossWinn 4.0Pre\cite{MossWinn4Manual}. Below $T_{\mathrm{N}}$, the M\"{o}ssbauer spectra can be modeled with a single value of hyperfine field (reported in Fig.~4f) and an increased linewidth with respect to the paramagnetic state. Alternatively, they may be described by a temperature independent linewidth and a distribution of hyperfine fields. Both approaches suggest that there is a distribution of $\textbf{\textit{H}}_{\textbf{hf}}$ at the Fe sites in the magnetic phase.

Nuclear magnetic resonance (NMR) measurements of CaK(Fe$_{0.951}$Ni$_{0.049}$)$_4$As$_4$ (total mass $\sim$$5$\,mg) were carried out on $^{75}$As ($I$\,$=$\,$\frac{3}{2}$, $\gamma/2\pi$\,$=$\,$7.2919$\,MHz/T, $Q$\,$=$\,$0.29$\,barns) by using a lab-built, phase-coherent, spin-echo pulse spectrometer. The spin echo was observed with a sequence of $\frac{\pi}{2}$ pulse ($2.2\,\mathrm{\mu}$s)\,-\,$30$\,$\mathrm{\mu}$s\,-\,$\pi$ pulse ($4.4\,\mathrm{\mu}$s). The $^{75}$As-NMR spectra were obtained at a fixed frequency, $f\,=\,43.2$\,MHz, with sweeping the magnetic field up to $8.25$\,T. The magnetic field was applied parallel or perpendicular to the crystallographic $c$-axis. $^{75}$As NMR spectra was simulated with two (for $\textbf{\textit{H}}\parallel\textbf{\textit{c}}$ above $T_{\mathrm{N}}$ and $\textbf{\textit{H}}\perp\textbf{\textit{c}}$) or three ($\textbf{\textit{H}}\parallel\textbf{\textit{c}}$ below $T_{\mathrm{N}}$) sets of peaks with distinct values of $\textbf{\textit{H}}_{\mathrm{hf}}$ and quadrapole interaction, $\nu_{\mathrm{Q}}$. The ratio of the integrated area of As1 to As2 is $1$:$1$ in all the spectra. The ratio of As1 with $-H_\mathrm{hf}$ to $+H_\mathrm{hf}$ is also $1$:$1$ in the spectra for $\textbf{\textit{H}}\parallel\textbf{\textit{c}}$ below $T_{\mathrm{N}}$.

The symmetry analysis was performed used the ISOTROPY software suite\cite{Campbell_2006_ISODISPLACE} to systematically analyze the possible magnetic orders of CaKFe$_{4}$As$_{4}$ with magnetic space group \textit{P4/mmm1'}. A detailed description of the analysis is presented in supplemental materials\cite{Supplement}.

In the first-principles calculations, the effect of electron doping on the magnetic properties of CaKFe$_4$As$_4$ was studied using the virtual crystal approximation (VCA) and the all-electron full-potential linearised augmented-plane wave basis (Elk FP-LAPW code).\cite{Elk} The structure parameters were fixed at the measured room-temperature values for undoped samples.\cite{Mou_2016_CaKFe4As4ARPES}  

\end{methods}

\begin{addendum}
 \item The authors would like to acknowledge W. Straszheim for his expertise and assistance with chemical analysis, D. S. Robinson for experimental support of the x-ray scattering study, and T. Kong for assistance during early stages of sample preparation. We also acknowledge discussions with E. Berg, A. Chubukov, and S. Kivelson. 
 
Work at the Ames Laboratory was supported by the Department of Energy, Basic Energy Sciences, Division of Materials Sciences \& Engineering, under Contract No. DE-AC02-07CH11358. WRM was supported by the Gordon and Betty Moore Foundation’s EPiQS Initiative through Grant GBMF4411. This research used resources of the Advanced Photon Source, a U.S. Department of Energy (DOE) Office of Science User Facility operated for the DOE Office of Science by Argonne National Laboratory under Contract No. DE-AC02-06CH11357. VB and RV are  supported by the Deutsche 
Forschungsgemeinschaft (DFG) under grant SFB/TRR49. RMF is supported by the Office of Basic Energy Sciences, U.S. Department of Energy, under award DE-SC0012336. PPO acknowledges support from Iowa State University Startup Funds. 

\item[Author contributions] PCC, SLB and WRM initiated this work; WRM grew the samples; WRM, QPD, AK, SLB, AS, KK, and AEB performed measurements and analyzed the data; VB, RV, PPO, CDB, and RMF performed the theoretical interpretations and computational work; WRM and AK performed the symmetry analysis; RV, RMF, AIG, YF, AEB, and PCC guided the work; WRM, AK, and AEB drafted the manuscript and all authors participated in the writing and review of the final draft.

\item[Data Availability] The data that support the findings of this study are available from
the corresponding author upon request.

\item[Supplemental Materials] Supplemental material referenced in the text can be found on-line.

\item[Competing Interests] The authors declare that they have no
competing financial interests.

\item[Correspondence] Correspondence and requests for materials should be addressed to P. C. Canfield~(email: canfield@ameslab.gov).

\end{addendum}

\pagebreak

\section{Supplemental materials}

\subsection{Composition analysis.}

Figure S1 presents the transition-metal ratio determined by wavelength-dispersive spectroscopy (WDS). In both the Co and Ni substitution series, the measured ratios were lower than the nominal ratios batched in the crucibles. Figure S2 presents the evolution of the $c$ lattice parameter with measured Co- and Ni-content. To determine $c$, X-ray diffraction was carried out on single crystals [cleaved along (001) planes] affixed to the sample holder with vacuum grease (Dow Corning High Vacuum Grease) in a Rigaku Miniflex II diffractometer (with a Cu tube and monochromator) using the method described in Jesche et al.\cite{Jesche2016_CrystalXRD}.

\begin{center}
	\includegraphics[width=8.6cm]{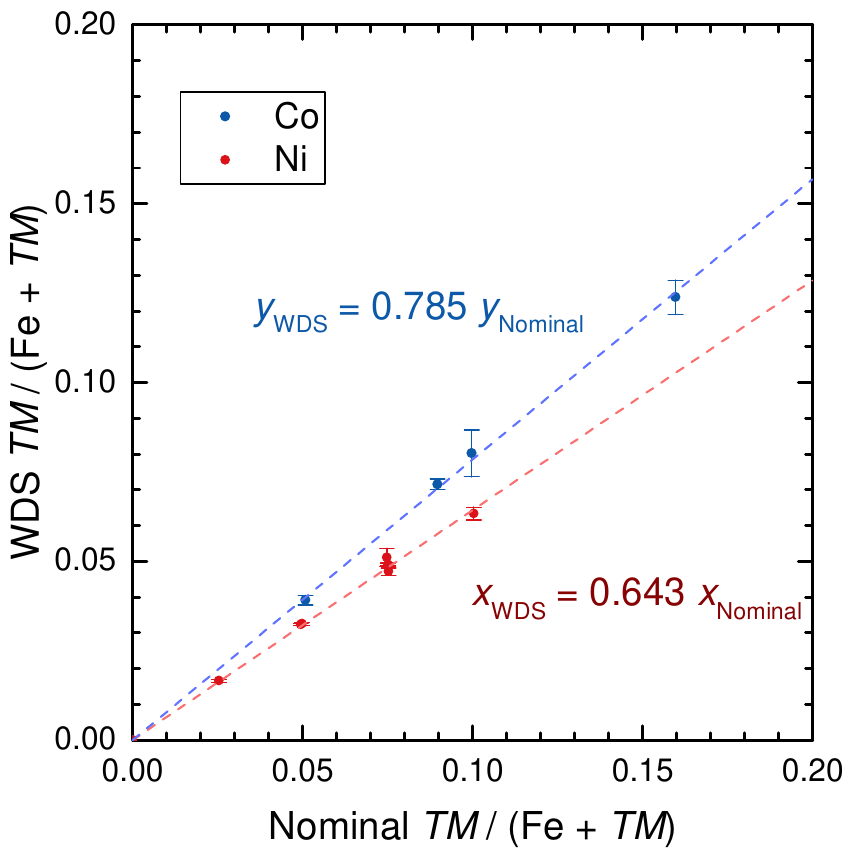}%
	\label{WDSvsBatched}
\end{center}
\noindent {\bf Figure S1. Measured transition-metal ratio vs nominal.} The transition-metal (\textit{TM}) atomic ratios [($x =$ Co/(Fe+Co) or $y =$ Ni/(Fe+Ni)] determined by wavelength-dispersive spectroscopy (WDS). Vertical error bars represent the standard deviation of 12 composition points for each sample.

\begin{center}
	\includegraphics[width=8.6cm]{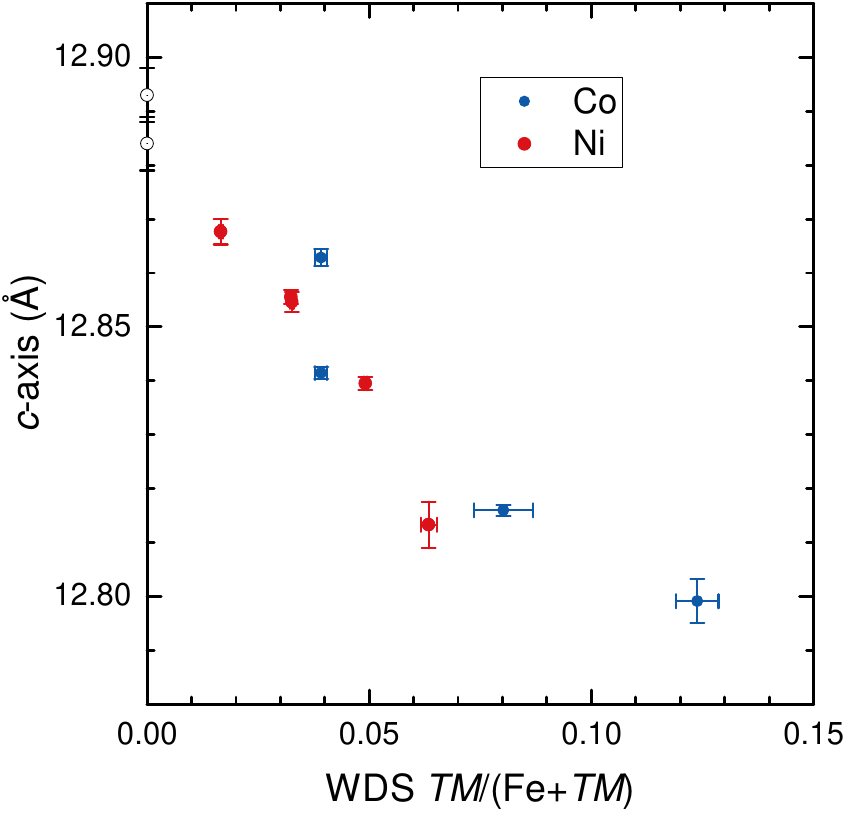}%
	\label{CvsDoping}
\end{center}
\noindent {\bf Figure S2. \textbf{\textit{c}} lattice parameter vs composition.} Substitution of Co and Ni into CaKFe$_{4}$As$_{4}$ reduces the $c$ lattice parameter. Horizontal error bars represent the standard deviation of 12 composition points for each sample. Vertical error bars denote the standard error from fitting the diffraction patterns.

\subsection{Theoretical analysis of NMR data: hyperfine fields for different magnetic orders}
Here we derive the internal fields experienced by the As atoms in different magnetic configurations. Since the unit cell is doubled in the double-Q magnetic phases, we consider four As atoms per Fe-As layer (see Fig.~S3). The theoretically calculated magnetic hyperfine field, $\textbf{\textit{H}}_{\text{hf}}$, at As-site $\text{As}_j$ with $j = 1,2,3,4$ is obtained from $\textbf{\textit{H}}_{\text{hf}}(\text{As}_j) = \sum_{i=1}^4 \mathcal{A}_{j,i} \textbf{\textit{m}}_i$ (Ref.~\onlinecite{Smerald2011_BaFe2As2NMR_T1}). Here, $\mathcal{A}^{\alpha \beta}_{j,i}$ ($\alpha,\beta = a,b,c$) is the nuclear-electron coupling tensor and the sum is over the Fe moments $\textbf{\textit{m}}_i$ of the four Fe-sites belonging to the same plaquette centered at the As site $\text{As}_j$ (see Fig.~S3). The matrix elements $\mathcal{A}_{j,i}$ are all related by symmetry such that only $\mathcal{A}_{1,1}^{\alpha, \beta}$ are independent parameters. Explicitly, denoting $\mathcal{A}_{1,1}^{\alpha, \beta} \equiv \mathcal{A}_1$ it holds that $\mathcal{A}_{1,2} \equiv \mathcal{A}_2 = \sigma_v(bc) \mathcal{A}_2 \sigma_v(bc)$, $\mathcal{A}_{1,3} \equiv \mathcal{A}_3 = C_2 \mathcal{A}_2 C_2$ and $\mathcal{A}_{1,4} \equiv \mathcal{A}_4 = \sigma_v(ac) \mathcal{A}_2 \sigma_v(ac)$. Here, $\sigma_v(bc) = \left( \begin{smallmatrix} -1 & & \\ & 1 & \\ &&1 \end{smallmatrix} \right)$ corresponds to mirror reflection with respect to the $bc$ plane, $\sigma_v(ac) = \left( \begin{smallmatrix} 1 & & \\ & -1 & \\ &&1 \end{smallmatrix} \right)$ corresponds to a mirror reflection with respect to the $ac$ plane, and $C_2(c) = \left( \begin{smallmatrix} -1 & & \\ & -1 & \\ &&1 \end{smallmatrix} \right)$ refers to a $\pi/2$ rotation with respect to the $c$-axis. The matrix elements $\mathcal{A}_{j,i}$ at the other As sites follow from symmetry considerations and are depicted in Fig.~S3. In fact, due to tetragonal symmetry only three independent parameters exist, \emph{e.g.}, $\mathcal{A}^{aa}_{1,1}, \mathcal{A}^{ac}_{1,1}$, $\mathcal{A}^{cc}_{1,1}$. 

Straightforward calculations yield the following hyperfine fields for different double-Q magnetic configurations. For the hedgehog SVC (see Fig.~1c), we find: $\textbf{\textit{H}}_{\text{hf}}(\text{As}_1) = -\textbf{\textit{H}}_{\text{hf}}(\text{As}_3) = (0,0,8 m \mathcal{A}^{ca})$, $\textbf{\textit{H}}_{\text{hf}}(\text{As}_2) = \textbf{\textit{H}}_{\text{hf}}(\text{As}_4) = (0,0,0)$, where $m$ is the size of the Fe moment in the $ab$-plane. In contrast, for loops SVC order (Fig.~1d) the hyperfine field vanishes at all As-sites. Spin-charge-density wave (SCDW) order (Fig.~1b) with moments along $\textbf{\textit{c}}$ yields $\textbf{\textit{H}}_{\text{hf}}(\text{As}_1) = 4 m_c \mathcal{A}^{ac} (1,1,0)$ and fields that are rotated by $-\frac{\pi}{2}$ with respect to each other from $a = 1$ to $a=4$.

For single-Q magnetic configurations with an arbitrary Fe moment $\left( m_a, \, m_b, \, m_c \right)$, we obtain the following results. N\'eel order gives $\textbf{\textit{H}}_{\text{hf}}(\text{As}_1) = - \textbf{\textit{H}}_{\text{hf}}(\text{As}_2) = \textbf{\textit{H}}_{\text{hf}}(\text{As}_3) = -\textbf{\textit{H}}_{\text{hf}}(\text{As}_4) = 4 \mathcal{A}^{aa} (m_b,m_a, 0)$. Stripe-type spin-density wave (SSDW) order (Fig.~1a) yields $\textbf{\textit{H}}_{\text{hf}}(\text{As}_1) = \textbf{\textit{H}}_{\text{hf}}(\text{As}_2) = -\textbf{\textit{H}}_{\text{hf}}(\text{As}_3) = -\textbf{\textit{H}}_{\text{hf}}(\text{As}_4) = 4 \mathcal{A}^{ac} (m_c,0, m_a)$ for $\left( \pi ,0 \right)$ wave-vector and $\textbf{\textit{H}}_{\text{hf}}(\text{As}_1) = 4 \mathcal{A}^{ac}(0,m_c, m_b)$ for $\left(0, \pi \right)$ wave-vector.

\begin{center}
	\includegraphics[width=8.6cm]{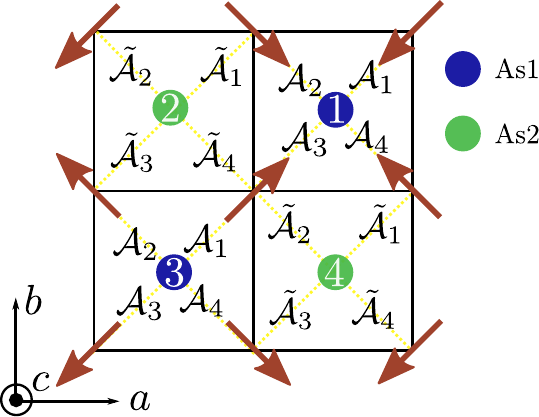}%
	\label{Hyperfine}
\end{center}
\noindent {\bf Figure S3. Definition and labeling of hyperfine tensors.} Definition and labeling of hyperfine tensors $\mathcal{A}_{j,i}$ within the magnetic unit cell, which contains four As atoms (As1 in blue, As2 in green) labeled by $j$. Each As interacts with the magnetic moments on its four Fe neighbors (brown arrows), which are labeled by $i$, via the hyperfine interaction tensor $\mathcal{A}_{j,i}$. The figure shows that $\mathcal{A}_{1,i} = \mathcal{A}_{3,i} \equiv \mathcal{A}_{i}$ and $\tilde{\mathcal{A}}_{2,i} = \tilde{\mathcal{A}}_{4,i} = \tilde{\mathcal{A}}_{i}$, where $\tilde{\mathcal{A}}_i$ is related to $\mathcal{A}_i$ by a mirror reflection about the Fe-plane.

\subsection{Symmetry analysis.}
A symmetry analysis was performed to systematically explore the possible magnetic orders of CaKFe$_{4}$As$_{4}$ [space group \textit{P4/mmm} (No. $123$)] and identify the ordered phase adopted by the Ni-doped compound. The CaKFe$_{4}$As$_{4}$ structure does not have atoms at the most general crystallographic sites\cite{Mou_2016_CaKFe4As4ARPES}. As a result, not all irreducible representations of the magnetic space group \textit{P4/mmm1'} allow atomic shifts, lattice distortions, charge asymmetries, or magnetic moments at occupied sites. Thus, they would not generate a phase transition and were not considered.

The experimental data was used to constrain the possible phases. M\"{o}ssbauer and NMR data indicate magnetic order, requiring that time-reversal symmetry is broken by the new phase. This allowed us to restrict our search to the time-odd magnetic irreducible representations. Only $96$ commensurate irreducible representations satisfy this requirement and develop finite magnetic moments at occupied atomic sites.

In addition, high-energy x-ray diffraction reveals no distortion of the diffraction peaks suggesting that the structure remains tetragonal. This suggests that only tetragonal subgroups of P4/mmm1' should be considered. The absence of new superstructure peaks below $T_{\mathrm{N}}$ lead us to look for irreducible representations which leave the chemical unit cell unchanged. It is possible for the geometry and periodicity of the chemical unit cell to remain unchanged while the magnetic order exhibits a larger period. These two constraints reduce the $96$ possible magnetic orders to $24$; $1$ \textit{c}-axis ferromagnet, $7$ antiferromagnetic structures with moments along \textit{c}-axis, and the $16$ modifications of spin-vortex crystals (SVCs) with moments in the \textit{ab}-plane, as described in the main text.

Heat capacity measurements of a  CaK(Fe$_{0.951}$Ni$_{0.049}$)$_{4}$As$_{4}$ sample (inset in Fig.~2a) reveal a second-order phase transition at $T_{\mathrm{N}}$. The temperature dependence of the hyperfine fields, $\textbf{\textit{H}}_{\mathrm{hf}}(T)$, at the Fe- and As1-sites (Fig.~4f) also supports this conclusion. At a second-order phase transition, only one order parameter may become critical and only one generator of the symmetry group may be broken.\cite{Landau1937_TheoryOfPhaseTransitions} This restriction requires that the new crystallographic phase has a space group that is a maximal subgroup of the parent space group. If we apply this criteria to the $96$ irreducible representations we generate the same list of $24$ magnetic structures obtained in the previous paragraph. Obtaining the same list from independent requirements strengthens our confidence in the result.

\begin{center}
	\includegraphics[width=8.6cm]{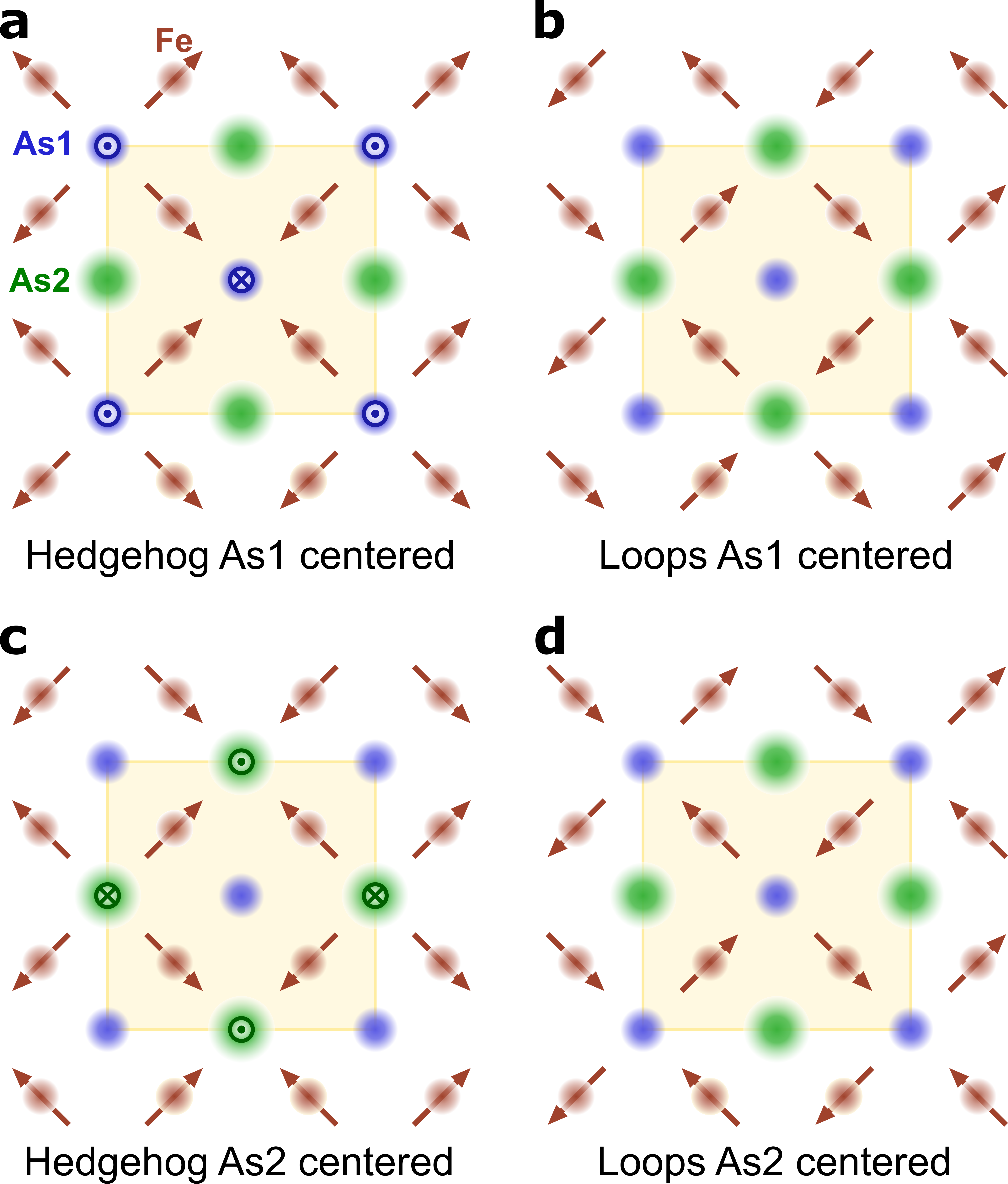}%
	\label{Centerings}
\end{center}
\noindent {\bf Figure S4. Spin-vortex crystal (SVC) motif centerings.} The hedgehog and loops variations of SVC order can be centered on either As1 or As2.

To further reduce the list, we return to the results of the local microscopic magnetic probes. The fits of the M\"{o}ssbauer spectra suggest that the moments at the Fe-sites are perpendicular to the crystallographic \textit{c}-axis. The eight \textit{c}-axis ferromagnetic and antiferromagnetic orders described above are inconsistent with this result. This leaves the 16 modifications of SVC order. 
For each of the two variations of SVC ("hedgehog" and "loops" in Figs. 1c,d) there are two inequivalent centerings. The motifs can be centered around As1 or As2 as depicted in Fig.~S4.
Only the hedgehog type centered on As1 (as depicted in Fig.~1c) generates an alternating $\textbf{\textit{H}}_{\mathrm{hf}}$\,$\parallel$\,$\textbf{\textit{c}}$ at As1 and $\textbf{\textit{H}}_{\mathrm{hf}}$\,$=$\,$\textbf{0}$ at As2. 

The hedgehog SVC motif can be stacked in four different ways without changing the chemical unit cell. (1) The moment motif is aligned between the iron layers (ferromagnetic stacking). (2) All the moments reverse every layer (antiferromagnetic stacking). (3 and 4) A double stacking of two aligned layers followed by two reversed layers can be realized without affecting the chemical unit-cell. There are two variations of the latter as the moment motif can either be reversed across the Ca plane or the K plane. Neutron diffraction could indicate which stacking is favored. Stackings 1 and 2 would give magnetic diffraction peaks at ($\frac{1}{2}$,$\frac{1}{2}$,0) in the CaKFe$_4$As$_4$ Brillouin zone. 

The tetragonal "C$_\mathrm{4}$" in the hole-doped ($Ae_{1-x}A_x$)Fe$_2$As$_2$ is reported to be a SCDW with $c$-axis moments like that shown in Fig.~1b. Surprisingly, a SCDW is orthorhombic in CaKFe$_4$As$_4$. In the $Ae$Fe$_2$As$_2$ structure all As sites are symmetry equivalent (green and blue As atoms would be equivalent in Fig.~1 and have the same distance from the iron plane). In $Ae$Fe$_2$As$_2$'s paramagnetic state \textit{4}-fold axes parallel to [001] are present through the As-sites and $\bar{\textit{4}}$ (roto-inversion) centers at the Fe- and $Ae$-sites. In the SCDW phase the \textit{4}-fold axes are broken by the inequivalent Fe-sites (one moment bearing, one not). The phase remains tetragonal because there are still $\bar{\textit{4}}$ or $\bar{\textit{4}}\textit{'}$ (time-reversal roto-inversion) centers at each Fe- and \textit{Ae} site which transform the As-sites above the plane to those below. On the other hand, in CaKFe$_4$As$_4$ these arsenics are not equivalent so there is no $\bar{\textit{4}}$ (roto-inversion) or $\bar{\textit{4}}\textit{'}$ symmetry. In this case the SCDW phase is orthorhombic (point group \textit{mmm}) because there are no other variations of \textit{4}-fold symmetry present. Nevertheless, orthorhombic distortions could be small.

Our identification of SVC order is strengthened by corroborating evidence from several techniques. 
Each of the phases other than SVC is inconsistent with at least two experimental observations. For example, the SCDW phase breaks more than one symmetry generator and, therefore, cannot be achieved through a second-order phase transition. In addition, it is inconsistent with the hyperfine field orientation at As observed in NMR and M\"{o}ssbauer.

\subsection{Incommensurate magnetic phases.}
Relaxing our requirement for an unchanged chemical unit cell we can consider the $102$ incommensurate irreducible representations of the CaKFe$_4$As$_4$ structure. $16$ of these produce tetragonal magnetic structures and only $8$ of those are consistent with a second-order phase transition. Interestingly, these $16$ are incommensurate variations of $c$-axis ferromagnetism and antiferromagnetism as well as $4$ SVC orders. As in the commensurate case, the first two groups are inconsistent with the orientation of $\textbf{\textit{H}}_{\mathrm{hf}}$ indicated by the M\"{o}ssbauer fits. As a result, even if weak superlattice peaks (indicating an incommensurate chemical displacements) were not resolved it is likely that the magnetic order is an incommensurate variety of SVC.

\subsection{Ginzburg-Landau theory of magnetic orders in the presence of a symmetry-breaking crystalline field}

Magnetic order with propagation vectors $\bfqq_1 = (\pi,0)$ and $\bfqq_2 = ( 0, \pi)$ is described by
\begin{equation}
\label{eq:1}
\bfm(\bfrr) = \bfmm_1 \cos (\bfqq_1 \cdot \bfrr) + \bfmm_2 \cos (\bfqq_2 \cdot \bfrr) \,,
\end{equation}
where $\bfrr$ denotes an Fe site in the lattice and $\bfmm_i$ are $N=3$ component magnetic order parameters. The homogeneous Ginzburg-Landau free energy is constrained by tetragonal and time-reversal symmetries to (see also Refs.~\onlinecite{Lorenzana_2008_CompetingOrdersInFeSCs,Eremin_2010_SDWsInFeSCs,Brydon_2011_GinzburLandauFeSCs,Cvetkovic_2013_SpGrpSymmAndSOHamltnFeSCers,Wang_2015_TetragMagOrderFeArsenides,Christensen_2015_SpinReorientation,Gastiasoro_2015_2QinFeSCs,Hoyer_2016_DisorderPromotedC4Phase,Fernandes_2016_VestigialOrders,OHalloran_2017_SVCinFeSCers}).
\begin{align}
\label{eq:2}
F &= \frac{a}{2} \bigl( M_1^2 + M_2^2\bigr) + \frac{u}{4} \bigl( M_1^2 + M_2\bigr)^2 - \frac{g}{4} \bigl( M_1^2 - M_2^2)^2 + w (\bfmm_1 \cdot \bfmm_2)^2\,,
\end{align}
where $M_i = |\bfmm_i|$, $a \propto T - T_N$ with mean-field transition temperature $T_N$ and quartic coefficients $u, g, w$. Although such a free energy can be derived from microscopic models\cite{Lorenzana_2008_CompetingOrdersInFeSCs,Eremin_2010_SDWsInFeSCs,Brydon_2011_GinzburLandauFeSCs,Cvetkovic_2013_SpGrpSymmAndSOHamltnFeSCers,Wang_2015_TetragMagOrderFeArsenides,Christensen_2015_SpinReorientation,Gastiasoro_2015_2QinFeSCs,Hoyer_2016_DisorderPromotedC4Phase,Fernandes_2016_VestigialOrders,OHalloran_2017_SVCinFeSCers}, here we focus on a purely phenomenological description, as the importance of our results does not depend on model details. The free energy in Eq.~\eqref{eq:2} is known to allow for three different types of magnetically ordered ground states (for $a<0$): (i) Stripe-type spin-density wave (SSDW) for which only one of the $M_i$ is non-zero; (ii) Spin-charge-density wave (SCDW) for which $\bfmm_1 = \pm \bfmm_2$; (iii) spin-vortex crystal (SVC) for which $|\bfmm_1| = |\bfmm_2|$ with non-collinear $\bfmm_1 \perp \bfmm_2$. 
For convenience, we write $M_i = M$ whenever it is non-zero such that the three magnetic states are explicitly given by $\bfm_{\text{SSDW}}(\bfrr) = M \hat{\bfm} \cos (\bfqq_i \cdot \bfrr)$, $\bfm_{\text{SCDW}}(\bfrr) = M \hat{\bfm} [ \cos (\bfqq_1 \cdot \bfrr) + \cos(\bfqq_2 \cdot \bfrr)]$ and $\bfm_{\text{SVC}}(\bfrr) = M [\hat{\bfm}_1 \cos (\bfqq_1 \cdot \bfrr) + \hat{\bfm}_2 \cos (\bfqq_2 \cdot \bfrr)]$. Here, $\hat{\bfm}_i =  \bfmm_i/|\bfmm_i|$ denotes the direction of the (non-zero) order parameter $\bfmm_i$. 

The magnetic ground state is determined by minimizing the free energy with respect to $M$ in the respective ordered phases, $\frac{\partial F}{\partial M} = 0$. We obtain $M_{\text{SSDW}} = \sqrt{\frac{-a}{u-g}}$, $M_{\text{SCDW}} = \sqrt{\frac{-a}{2(u+w)}}$ and $M_{\text{SVC}} = \sqrt{\frac{-a}{2u}}$. Inserting these expressions into the free energy yields the different ground state energies $E_{\text{SSDW}} = -\frac{a^2}{4(u - g)}$, $E_{\text{SCDW}} = - \frac{a^2}{4 (u + w)}$ and $E_{\text{SVC}} = - \frac{a^2}{4 u}$. The ratios of ground state energies are thus $\frac{E_{\text{SVC}}}{E_{\text{SSDW}}} = \frac{u-g}{u}$ and $\frac{E_{\text{SCDW}}}{E_{\text{SSDW}}} = \frac{u-g}{u+w}$. It then follows that SVC is the ground state for $-g, w > 0$, while SSDW is the lowest energy state for $g > \text{max} \{0, -w \}$ and SCDW for $w < \text{min}\{0,-g\}$, in agreement with previous analyses\cite{Lorenzana_2008_CompetingOrdersInFeSCs,Eremin_2010_SDWsInFeSCs,Brydon_2011_GinzburLandauFeSCs,Cvetkovic_2013_SpGrpSymmAndSOHamltnFeSCers,Wang_2015_TetragMagOrderFeArsenides,Christensen_2015_SpinReorientation,Gastiasoro_2015_2QinFeSCs,Hoyer_2016_DisorderPromotedC4Phase,Fernandes_2016_VestigialOrders,OHalloran_2017_SVCinFeSCers}.

The new ingredient in the case of the \textit{AeA}Fe$_{4}$As$_{4}$ compounds is the difference in the heights of the two As atoms staggered above and below the Fe plane, which effectively breaks the glide plane symmetry of the Fe-As layers. The symmetry breaking field generated by this crystal structure can be parametrized in terms of a vector $\boldsymbol{\eta} = \eta \, \hat{\mathbf{z}}$ that couples linearly to the vector spin chirality according to:
\begin{align}
\label{eq:3}
F_\eta &= - \frac12 \boldsymbol{\eta} \cdot (\bfmm_1 \times \bfmm_2) \,.
\end{align}
The emergence of this term can be derived from symmetry considerations. As discussed in details in Ref.~\onlinecite{Fernandes_2016_VestigialOrders}, inside the SVC phase, besides long-range magnetic order, a composite order parameter also condenses: $\bfmm_1 \times \bfmm_2$, called spin-vorticity-density wave. This "chiral" order parameter, which lives on the plaquettes of the Fe-As layer, is related to the spin-vortices that appear in the SVC phase; it preserves time-reversal symmetry and has wave-vector $\bfqq_1 + \bfqq_2 = (\pi, \pi)$ in the Fe-only square lattice. Importantly, as shown in Ref.~\onlinecite{Fernandes_2016_VestigialOrders}, this order parameter has the same symmetry as a spin-current-density wave with the same propagation vector (also known as a triplet d-density wave), manifested as staggered spin loop currents along the plaquettes (see Fig. 1f of the main text). While charge loop currents generate magnetic fields, spin loop currents generate electric fields\cite{Sun_2004_SpinCurrentEField}. In particular, the staggered configuration of spin loop currents creates staggered electric fields of opposite directions above and below the Fe-plane. Because the As atoms located above and below the Fe-plane are themselves staggered, all As atoms either experience an electric field pointing up or down (see Fig. 1f of the main text). Consequently, the two As atoms, As1 and As1, become inequivalent, and their heights with respect to the Fe plane also become different.

Thus, the symmetry breaking field induced by the crystal, $\eta$, triggers immediately a spin-vorticity density-wave -- similarly to how uniaxial strain triggers nematic order. Although the spin-vorticity density-wave does not trigger magnetic order, the fact that this order parameter is only non-zero inside the SVC phase implies that, for large enough $|\eta| > \eta_c$, the magnetic ground state necessarily becomes SVC. Indeed, including $F_\eta$ does not change the size of $M_i$ and the energies $E_\alpha$ for SSDW and SCDW orders since the spin vorticity vanishes for these two phases. In contrast, $F_\eta$ increases the size of $M_i$ in the SVC phase to $M_{\text{SVC}} = \sqrt{\frac{-(2a- \eta)}{4u}}$ and changes its energy to $E_{\text{SVC}} = - \frac{(a - \frac{\eta}{2})^2}{4 u}$. The critical value $\eta_c$ that enforces an SVC ground state in the case that $E_{\text{SVC}}(\eta = 0) > E_{\text{GS}} \equiv \text{min} \{E_{\text{SSDW}}, E_{\text{SCDW}} \}$ follows from a straightforward calculation: 
\begin{align}
\label{eq:4}
\frac{\eta_c}{2 |a|} = \sqrt{\frac{E_{\text{GS}}}{E_{\text{SVC}}(0)}} - 1 \,.
\end{align}

Therefore, the closer the energy of the SVC state is to the energy of the magnetic ground state (in the absence of the symmetry breaking field), the smaller the value of the symmetry breaking field required to make the SVC state the ground state.

\subsection{\textit{ab initio.}}
In the virtual crystal approximation (VCA) method, homogeneous doping is imposed by replacing Fe with an artificial nucleus of a fractional charge between that of Fe and Co. The $\textbf{\textit{k}}$-point mesh was set to $(5\times 5\times 5)$ dimensions. For each dopant concentration, we study the relativistic total energies of two magnetic orders imposed on the Fe sublattice, namely, the stripe order along the $a$-axis and the ``hedgehog'' SVC order with Fe spins lying in-plane and centered on As-sites near the K-plane, as pointed out in the main text.
These two configurations were suggested to represent the main competing fluctuations in the undoped compound, based on the NMR measurements and \textit{ab initio} findings\cite{Cui2017_CaKFe4As4NMR}. In the present calculations, the structure remains fixed and we consider the pure effect of electron count on the magnetic energies. Since the generalized-gradient approximation (GGA) overestimates the magnitude of Fe moments in iron pnictides, we use the recently proposed reduced Stoner theory (RST)\cite{Ortenzi2012_AccountingForSpinFluctuations,Ortenzi2015_AnomalousTempDepLocalMomentsCaFe2As2} to account for the effect of fluctuation-induced renormalization of Fe moments and match their value with the experimental estimate for the doped compound ($\sim 0.2\,\mu_\mathrm{B}$).

Our \textit{ab initio} findings from the VCA+RST simulations indicate that the energy separation between different magnetic states is very sensitive to the size of Fe moments. When the renormalized Fe moments are close to the experimental estimate, the total energies of the two selected orders (stripe vs hedgehog shown in Fig.~S5) are almost degenerate in the undoped material. This suggests that fluctuations between these two configurations suppress magnetism in the undoped CaKFe$_4$As$_4$. Upon adding $0.05$ electrons per transition-metal (\textit{TM}), the hedgehog configuration prevails over the stripe order by almost $34$~meV/\textit{TM}. Higher electron additions ($\geq$\,$0.10$ electrons/\textit{TM}) leads again to the degeneracy of both types of order.

\begin{center}
	\includegraphics[width=8.6cm]{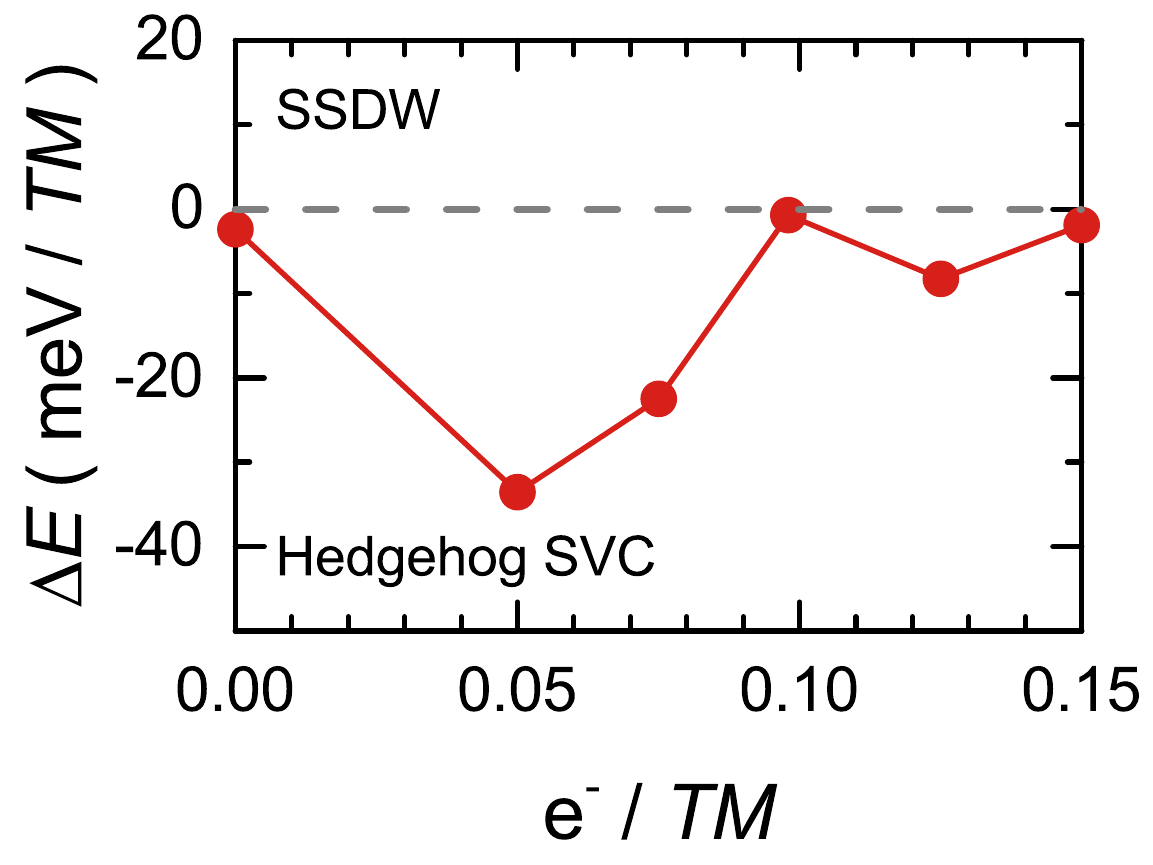}%
	\label{AbInitio}
\end{center}
\noindent {\bf Figure S5. Magnetic structure vs additional electron.} Energy difference between the "hedgehog" spin-vortex crystal and stripe-type phases of CaKFe$_4$As$_4$ vs electron doping per transition-metal [e$^-$/\textit{TM} $\approx$\,$y$\,$=$\,$\mathrm{Co}/(\mathrm{Fe}+\mathrm{Co})$]. Fluctuations of Fe magnetism are simulated using the reduced Stoner theory with $r=0.75$.

\end{document}